\documentstyle[12pt,epsfig,twoside]{article}
\epsfverbosetrue
\newcommand{\beq}{\begin{equation}}
\newcommand{\eeq}{\end{equation}}
\newcommand{\beqn}{\begin{eqnarray}}
\newcommand{\eeqn}{\end{eqnarray}}
\newcommand{\beqns}{\begin{eqnarray*}}
\newcommand{\eeqns}{\end{eqnarray*}}
\newcommand{\vs}{\\[0.3cm]\indent}

\newcommand{\hm}{\hspace{-0.05cm}}

\newcommand{\intl}{\int\limits}
\newcommand{\ointl}{\oint\limits}
\newcommand{\mc}{\multicolumn}
\newcommand{\e}{\epsilon}

\newcommand{\MSbm}{{\overline{\rm MS}}}
\def\NP{{\it Nucl. Phys.}}
\def\PL{{\it Phys. Lett.}}
\def\PR{{\it Phys. Rev.}}
\def\PRep{{\it Phys. Rep.}}
\def\PRL{{\it Phys. Rev. Lett.}}

\def\ZP{{\it Z. Phys.}}
\def\TAU{Talk given at the TAU'96 Conference, Colorado, 1996}

\def\ea{{\it et al.}}
\def\Cl{Collaboration}
\def\MeVE{~MeV}
\def\pc{$\%$}

\def\asZ{$\alpha_s(M_{\rm Z}^2)$}

\def\assz{$\alpha_s(s_0)$}
\def\ee{$e^+e^-$}

\def\pms{$\,\pm\,$}
\def\aqed{$\alpha(s)$}
\def\aqedZ{$\alpha(M_{\rm Z}^2)$}
\def\daqed{$\Delta\alpha(s)$}

\def\daqedhZ{$\Delta\alpha_{\rm had}(M_{\rm Z}^2)$}
\def\amuhad{$a_\mu^{\rm had}$}
\def\FOPTCI{$\rm FOPT_{\rm CI}$}
\def\ie{{\it i.e.}} 
 
\def\via{via} 

\def\rs{\raisebox{1.5ex}[-1.5ex]}

\begin{document}

\begin{titlepage}
\setcounter{page}{1}

\begin{flushright} 
{\bf LAL 97-85}\\
{November 1997}
\end{flushright} 

\begin{center}
\vspace{2cm}
{\Large
  {\LARGE I}MPROVED {\LARGE D}ETERMINATION OF $\alpha(M_{\rm Z}^2$) AND THE\\[0.3cm]
  {\LARGE A}NOMALOUS {\LARGE M}AGNETIC {\LARGE M}OMENT OF THE {\LARGE M}UON \\
}
\vspace{1.6cm}
\begin{large}
Michel Davier\footnote{E-mail: davier@lal.in2p3.fr}
and Andreas H\"ocker\footnote{E-mail: hoecker@lalcls.in2p3.fr} \\
\end{large}
\vspace{0.5cm}
{\small \em $^{\rm b}$Laboratoire de l'Acc\'el\'erateur Lin\'eaire,\\
IN2P3-CNRS et Universit\'e de Paris-Sud, F-91405 Orsay, France}\\
\vspace{3cm}

{\small{\bf Abstract}}
\end{center}
{\small
\vspace{-0.2cm}
We reevaluate the hadronic contribution to the running of the QED 
f\/ine structure constant \aqed\ at $s=M_{\rm Z}^2$. We use data 
from \ee\ annihilation and $\tau$ decays at low energy and at the
$q\bar{q}$ thresholds, where resonances occur. Using so-called spectral 
moments and the Operator Product Expansion (OPE), it is shown that a 
reliable theoretical prediction of the hadronic production rate $R(s)$
is available at relatively low energies. Its application improves 
signif\/icantly the precision on the hadronic vacuum polarization 
contribution. We obtain \daqedhZ\,$=(277.8\pm2.6)\times10^{-4}$ yielding
$\alpha^{-1}(M_{\rm Z}^2)=128.923\pm0.036$. Inserting this value in a 
global electroweak f\/it using current experimental input, we 
constrain the mass of the Standard Model Higgs boson to be 
$M_{\rm Higgs}=129^{+103}_{-62}~{\rm GeV}$. Analogously,
we improve the precision of the hadronic contribution to the 
anomalous magnetic moment of the muon for which we obtain
$a_\mu^{\rm had}=(695.1\pm7.5)\times 10^{-10}$.

\noindent
}
\vspace{2.5cm}
\vfill
\centerline{\it (Submitted to Physics Letters B)}
\vspace{1cm}
\thispagestyle{empty}

\end{titlepage}

\newpage\thispagestyle{empty}{\tiny.}\newpage
\setcounter{page}{1}

%
% --------------------- Introduction ---------------------------------
%
\section{ Introduction }

The running of the QED f\/ine structure constant $\alpha(s)$
and the anomalous magnetic moment of the muon are famous observables
whose theoretical precisions are limited by second order loop ef\/fects
from hadronic vacuum polarization. Both magnitudes are related \via\
dispersion relations to the hadronic production rate in \ee\
annihilation,
\beq
\label{eq_rsigma}
      R(s) = \frac{\sigma_{\rm tot}(e^+e^-\!\rightarrow {\rm hadrons})}
                  {\sigma_0(e^+e^-\!\rightarrow \mu^+\mu^-)}
           = \frac{3s}{4\pi\alpha^2}
             \sigma_{\rm tot}(e^+e^-\!\rightarrow {\rm hadrons})~.
\eeq
While far from quark thresholds and at suf\/f\/iciently high energy
$\sqrt{s}$, $R(s)$ can be predicted by perturbative QCD, theory may 
fail when resonances occur, \ie, local quark-hadron duality is broken. 
Fortunately, one can circumvent this drawback by using \ee\ 
annihilation data of $R(s)$ and, as recently proposed in 
Ref.~\cite{g_2pap}, hadronic $\tau$ decays benef\/itting from the 
largely conserved vector current (CVC).
\vs
There is a strong interest in the electroweak phenomenology to
reduce the uncertainty in \aqedZ\ which at present is a serious limit
to further progress in the determination of the Higgs mass from 
radiative corrections in the Standard Model. The most constraining
observable so far has been ${\rm sin}^2\Theta_W$ obtained from
leptonic asymmetries at the Z pole with an achieved precision of 
$(\Delta {\rm sin}^2\Theta_W)_{\rm exp}=0.00022$~\cite{jerusalem}.
The uncertainty from the currently used \aqedZ\ value translates 
into $(\Delta {\rm sin}^2\Theta_W)_\alpha=0.00023$~\cite{blondel},
justifying the present work.
\vs
In this paper, we extend the use of the theoretical QCD prediction 
of $R(s)$ to energies down to 1.8~GeV. The reliability of this approach 
is justified by applying the Wilson Operator Product Expansion 
(OPE)~\cite{wilson} (also called SVZ approach~{\cite{svz}) and 
f\/itting the dominant nonperturbative power terms directly to the 
data by means of spectral moments. An analogous approach has 
been successfully applied to the theoretical prediction of the $\tau$ 
hadronic width, $R_\tau$, at $M_\tau\simeq1.8~{\rm GeV}$ in order to 
measure the strong coupling constant 
$\alpha_s(M_\tau^2)$~\cite{pichledib,aleph_as,cleo_as,aleph_asn}.
\vs
The paper is organized as follows: first a brief overview over the 
formulae used is given, then the spectral moments are def\/ined and
evaluated experimentally, and the corresponding theoretical evaluation 
is f\/itted. On this basis, theoretical predictions of the hadronic 
vacuum polarization contribution to the running of $\alpha(M_{\rm Z}^2)$ 
and to the anomalous magnetic moment of the muon, $(g-2)_\mu$,
from various energy regimes are determined and, with the addition of 
experimental data, f\/inal values for $\alpha(M_{\rm Z}^2)$ and
$a_\mu\equiv(g-2)_\mu/2$ are determined.

%
% ---------------- The running alpha  ---------------------------------
%
\section{ Running of the QED F\/ine Structure Constant}

The running of the electromagnetic f\/ine structure constant \aqed\
is governed by the renormalized vacuum polarization function,
$\Pi_\gamma(s)$. For the spin 1 photon, $\Pi_\gamma(s)$ is given 
by the Fourier transform of the time-ordered product of the 
electromagnetic currents $j_{\rm em}^\mu(s)$ in the vacuum 
$(q^\mu q^\nu-q^2g^{\mu\nu})\,\Pi_\gamma(q^2)=
i\int d^4x\,e^{iqx}\langle 0|T(j_{\rm em}^\mu(x)j_{\rm em}^\nu(0))|0\rangle$.  
With $\Delta\alpha(s)-4\pi\alpha\,{\rm Re}
\left[\Pi_\gamma(s)-\Pi_\gamma(0)\right]$, one has
\beq
    \alpha(s) \:=\: \frac{\alpha(0)}{1-\Delta\alpha(s)}~,
\eeq
where $4\pi\alpha(0)$ is the square of the electron charge in the 
long-wavelength Thomson limit. The contribution \daqed\ can naturally 
be subdivided in a leptonic and a hadronic part. 
\vs
The leading order leptonic contribution is given by
\beq
   \Delta\alpha_{\rm lep}(M_{\rm Z}^2)=
        \frac{\alpha(0)}{3\pi}\sum_\ell
            \left[ {\rm ln}\frac{M_{\rm Z}^2}{m_\ell^2}-\frac{5}{3}
                   + {\cal O}\left( \frac{m_\ell^2}{M_{\rm Z}^2} \right)
            \right]
             \:=\: 314.2\times10^{-4}~. 
\eeq
Using analyticity and unitarity, the dispersion integral for the 
contribution from the light quark hadronic vacuum polarization 
\daqedhZ\ reads~\cite{cabbibo}
\beq\label{eq_integral1}
    \Delta\alpha_{\rm had}(M_{\rm Z}^2) \:=\:
        -\frac{M_{\rm Z}^2}{4\pi^2\,\alpha}\,
         {\rm Re}\intl_{4m_\pi^2}^{\infty}ds\,
            \frac{\sigma_{\rm had}(s)}{s-M_{\rm Z}^2-i\epsilon}~,
\eeq
where 
$\sigma_{\rm had}(s)=16\pi^2\alpha^2(s)/s\cdot{\rm Im}\Pi_\gamma(s)$
is given by the optical theorem, and Im$\Pi_\gamma$ stands for the 
absorptive part of the hadronic vacuum polarization correlator. 
Through Eq.~(\ref{eq_rsigma}), the above dispersion relation can be
expressed as a function of $R(s)$:
\beq\label{eq_integral2}
    \Delta\alpha_{\rm had}(M_{\rm Z}^2) \:=\:
        -\frac{\pi\alpha M_{\rm Z}^2}{3}\,
         {\rm Re}\intl_{4m_\pi^2}^{\infty}ds\,
            \frac{R(s)}{s(s-M_{\rm Z}^2)-i\epsilon}~.
\eeq
Employing the identity $1/(x^\prime-x-i\epsilon)_{\epsilon\rightarrow0}
={\rm P}\{1/(x^\prime-x)\}+i\pi\delta(x^\prime-x)$, the 
integrals~(\ref{eq_integral1}) and (\ref{eq_integral2}) are
evaluated using the principle value integration technique.

%
% ------------------- Muon Magnetic Anomaly ------------------------------------
%
\section{Muon Magnetic Anomaly}

It is convenient to separate the prediction $a_\mu^{\mathrm SM}$ 
from the Standard Model into its dif\/ferent contributions
\beq
    a_\mu^{\mathrm SM} \:=\: a_\mu^{\mathrm QED} + a_\mu^{\mathrm had} +
                             a_\mu^{\mathrm weak}~,
\eeq
where $a_\mu^{\mathrm QED}=(11\,658\,470.6\,\pm\,0.2)\times10^{-10}$ is 
the pure electromagnetic contribution (see~\cite{krause1} and references 
therein), \amuhad\ is the contribution from hadronic vacuum polarization,
and $a_\mu^{\mathrm weak}=(15.1\,\pm\,0.4)\times10^{-10}
$~\cite{krause1,peris,weinberg} accounts for corrections due to the 
exchange of the weak interacting bosons up to two loops.
\vs
Equivalently to \daqedhZ, by virtue of the analyticity of the 
vacuum polarization correlator, the contribution of the hadronic 
vacuum polarization to $a_\mu$ can be calculated \via\ the dispersion 
integral~\cite{rafael}
\beq\label{eq_integralamu}
    a_\mu^{\mathrm had} \:=\: 
           \frac{\alpha^2(0)}{3\pi^3}
           \intl_{4m_\pi^2}^\infty ds\,\frac{R(s)\,K(s)}{s}~.
\eeq
Here $K(s)$ denotes the QED kernel~\cite{rafael2}
\beq
      K(s) \:=\: x^2\left(1-\frac{x^2}{2}\right) \,+\,
                 (1+x)^2\left(1+\frac{1}{x^2}\right)
                      \left({\mathrm ln}(1+x)-x+\frac{x^2}{2}\right) \,+\,
                 \frac{(1+x)}{(1-x)}x^2\,{\mathrm ln}x
\eeq
with $x=(1-\beta_\mu)/(1+\beta_\mu)$ and $\beta=(1-4m_\mu^2/s)^{1/2}$.
The function $K(s)$ decreases monotonically with increasing $s$. It gives
a strong weight to the low energy part of the integral~(\ref{eq_integral1}).
About 91\pc\ of the total contribution to \amuhad\ is accumulated at c.m. 
energies $\sqrt{s}$ below 2.1~GeV while 72\pc\ of \amuhad\ is covered by 
the two-pion f\/inal state which is dominated by the $\rho(770)$ 
resonance. Data from vector hadronic $\tau$ decays published by the 
ALEPH Collaboration provide a very precise spectrum of the two-pion 
f\/inal state as well as new input for the lesser known four-pion 
f\/inal states. This new information improves signif\/icantly
the precision of the \amuhad\ determination~\cite{g_2pap}.

%
% -------------------------- Theoretical prediction ---------------------
%
\section{ Theoretical Prediction of $R(s)$}

The optical theorem relates the total hadronic width at a given
energy-squared $s_0$ to the absorptive part of the photon vacuum 
polarization correlator
\beq
\label{eq_rimpi}
     R(s_0) = 12\pi{\rm Im}\Pi(s_0+i\epsilon)~.
\eeq
Perturbative QCD predictions up to next-to-next-to leading order
$\alpha_s^3$ are available for the Adler $D$-function~\cite{adler}
which is the logarithmic derivative of the correlator $\Pi$,
carrying all physical information:
\beq
\label{eq_adler}
     D(s) = - 12 \pi^2 s\frac{d\Pi(s)}{ds}~.
\eeq
This yields the relation
\beq
\label{eq_radler}
     R(s_0) = \frac{1}{2\pi i}
              \hm\ointl_{|s|=s_0}\hm\hm\frac{ds}{s} D(s)~,
\eeq
where the contour integral runs counter-clockwise around the 
circle from $s=s_0-i\e$ to $s=s_0+i\e$.
Choosing the renormalization scale to be the physical scale $s$,
additional logarithms in the perturbative expansion of $D$ are 
absorbed into the running coupling constant $\alpha_s(s)$. The
(massless) NNLO perturbative prediction of $D$ reads 
then~\cite{3loop}
\beq
\label{eq_pertd}
     D_{\rm P}(-s) = N_C\sum_f Q_f^2
            \left[ 1 + d_0\frac{\alpha_s(s)}{\pi}
                     + d_1\left(\frac{\alpha_s(s)}{\pi}\right)^{\!\!2}
                     + \tilde{d}_2\left(\frac{\alpha_s(s)}{\pi}\right)^{\!\!3}
                     + {\cal O}\left(\alpha_s^4(s)\right)
            \right]~,
\eeq
where $N_C=3$ for $SU(3)_C$ and $Q_f$ is the charge of the quark 
$f$. The coef\/f\/icients are $d_0=1$, $d_1=1.9857 - 0.1153\,n_f$,
$\tilde{d}_2=d_2+\beta_0^2\pi^2/48$, $\beta_0=11-2n_f/3$ and 
$d_2=-6.6368-1.2001\,n_f-0.0052\,n_f^2-1.2395\,(\sum_f Q_f)^2/N_C\sum_f Q_f^2$ 
with $n_f$ the number of involved quark flavours\footnote
{
   The negative energy-squared in $D_{\rm P}(-s)$ of Eq.~(\ref{eq_pertd})
   is introduced when continuing the Adler function from the spacelike
   Euclidean space, where it was originally def\/ined, to the timelike
   Minkowski space by virtue of its analyticity property.
}. 
The running of the strong coupling constant $\alpha_s(s)$ is governed 
by the renormalization group equation (RGE), known precisely to 
four-loop level~\cite{rit}.
\vs
Using the above formalism, $R(s_0)$ is easily obtained by evaluating
numerically the contour-integral~(\ref{eq_radler}). The solution 
is called contour-improved fixed-order perturbation 
theory (\FOPTCI) in the following. Another approach, usually chosen, 
is to expand $\alpha_s(s)$ in Eq.~(\ref{eq_pertd}) in powers of 
$\alpha_s(s_0)$ with coef\/f\/icients that are polynomials in 
${\rm ln(s/s_0)}$:
\beqn
\label{eq_alphasexp}
     \frac{\alpha_s(s)}{\pi} 
  &=&
      \frac{\alpha_s(s_0)}{\pi} \,-\, 
      \frac{1}{4}\beta_0\,{\rm ln}\frac{s}{s_0}
            \left(\frac{\alpha_s(s)}{\pi}\right)^{\!\!2}  \,-\,
      \left(\frac{1}{8}\beta_1\,{\rm ln}\frac{s}{s_0} 
            - \frac{1}{16}\beta_0^2\,{\rm ln}^2\frac{s}{s_0}
      \right)
            \left(\frac{\alpha_s(s)}{\pi}\right)^{\!\!3} \nonumber\\
  & & \,-\,
      \left(\frac{1}{128}\beta_2\,{\rm ln}\frac{s}{s_0} 
            - \frac{5}{64}\beta_0\beta_1\,{\rm ln}^2\frac{s}{s_0}
            + \frac{1}{64}\beta_0^3\,{\rm ln}^3\frac{s}{s_0}\right)
            \left(\frac{\alpha_s(s)}{\pi}\right)^{\!\!4} \,+\,\dots~.
\eeqn
Inserting the above series with the $D_{\rm P}$-function in Eq~(\ref{eq_radler}) 
and keeping terms up to oder $\alpha_s^3$ leads to the expression
\beq
\label{eq_r}
     R(s_0) = N_C\sum_f Q_f^2
            \left[ 1 + d_0\frac{\alpha_s(s_0)}{\pi}
                     + d_1\left(\frac{\alpha_s(s_0)}{\pi}\right)^{\!\!2}
                     + d_2\left(\frac{\alpha_s(s_0)}{\pi}\right)^{\!\!3}
                     + {\cal O}\left(\alpha_s^4(s_0)\right)
            \right]~,
\eeq
where the dif\/ference to $D_{\rm P}$ is of order $\alpha_s^3$ only. The 
solution~(\ref{eq_r}) will be referred to as fixed-order perturbation
theory (FOPT). There is an intrinsic ambiguity between FOPT and \FOPTCI. 
The numerical solution of the contour-integral~(\ref{eq_radler}) involves 
the complete (known) RGE and provides thus a resummation of all known 
higher order logarithmic terms of the expansion~(\ref{eq_alphasexp}) 
(see Ref.~\cite{pert} for comparison). Unfortunately, it is unclear if 
the resummation does not give rise to a bias of the f\/inal result.

%
% -------------------------- Mass corrections ---------------------------
%
\subsection*{\it Quark Mass Corrections}

Quark mass corrections in leading order are suppressed as 
$\sim m_q^2(s)/s$, \ie, they are small suf\/f\/iciently far away 
from threshold. Complete formulae for the perturbative prediction 
containing quark masses are provided in Refs.~\cite{broadhurst,kuhn1,kuhn2}
for the correlator $\Pi(s)$ and $R(s)$ up to order $\alpha_s$ exactly 
and numerically using Pad\'e approximants to order $\alpha_s^2$. We 
will learn from the numerical analysis that it suf\/f\/ices for the 
required level of accuracy to use the following expansion as an 
additive correction to the Adler $D$-function~\cite{kuhn3}
\beq
\label{eq_dmass}
      D_{\rm mass}(-s) 
         = - N_C \sum_f Q_f^2\frac{m_f^2(s)}{s}
                \left( 6 + 28\,\frac{\alpha_s(s)}{\pi}
                        + (294.8-12.3\,n_f)
                          \left(\frac{\alpha_s(s)}{\pi}\right)^{\!\!2}
                 \right)~.
\eeq
The running of the quark mass $m_f(s)$ is obtained from the 
renormalization group and is known to four-loop level~\cite{ritmass}.

%
% -------------------------- Nonperturbative corrections --------------
%

When using perturbative QCD at low energy scales one has to worry 
whether contributions from nonperturbative QCD could give rise to 
large corrections. The break down of asymptotic freedom is signalled 
by the emergence of power corrections due to nonperturbative 
ef\/fects in the QCD vacuum. These are introduced \via\ 
non-vanishing vacuum expectation values originating from quark 
and gluon condensation. It is convenient to use the Operator 
Product Expansion (OPE)~\cite{svz,reinders,bnp} in low energy
regions (or near quark thresholds), where nonperturbative 
ef\/fects come into play. One thus def\/ines
\beq
\label{eq_ope}
   D(-s) = \sum_{D=0,2,4,...} \frac{1}{(-s)^{D/2}}
                        \sum_{{\rm dim}{\cal O}=D} C_D(s,\mu)
                         \langle{\cal O_D}(\mu)\rangle~,
\eeq
where the Wilson coef\/f\/icients $C(s,\mu)$ include short-distance 
ef\/fects and the operator $\langle{\cal O_D}(\mu)\rangle$
collect the long-distance, nonperturbative dynamical 
information at the arbitrary separation scale $\mu$. The 
operator of Dimension $D=0$ ($D=2$) is the perturbative
prediction (with mass correction). The $D=4$ operator are linked
to the gluon and quark condensates. Ef\/fective approaches 
together with the vacuum saturation assumption are used to
compact the large number of dynamical $D=6$ ($D=8$) operator 
into one phenomenological operator $\langle{\cal O}_6\rangle$ 
($\langle{\cal O}_8\rangle$). The nonperturbative addition
to the Adler function reads then~\cite{bnp}
\beqn
\label{eq_nonpdef}
     D_{\rm NP}(-s) &=& N_C \sum_f Q_f^2
      \Bigg\{
          \frac{2\pi^2}{3}\left(1 - \frac{11}{18}\frac{\alpha_s(s)}{\pi}
                         \right)\frac{\left\langle\frac{\alpha_s}{\pi} 
                                           GG\right\rangle}{s^2} 
      \nonumber \\
     & &\hspace{2.2cm} 
         + \;8\pi^2\left(1 - \frac{\alpha_s(s)}{\pi}
                \right)\frac{\langle m_f\bar{q_f}q_f\rangle}{s^2}
         +  \frac{32\pi^2}{27}\frac{\alpha_s(s)}{\pi}
            \sum_k\frac{\langle m_k\bar{q_k}q_k\rangle}{s^2}
      \nonumber \\
     & &\hspace{2.2cm} 
          + \;12\pi^2\frac{\langle{\cal O}_6\rangle}{s^3}
          \;+\; 16\pi^2\frac{\langle{\cal O}_8\rangle}{s^4}
      \Bigg\}~,
\eeqn
with the gluon condensate, 
$\langle(\alpha_s/\pi) GG\rangle$, and the quark condensates,
$\langle m_f\bar{q_f}q_f\rangle$. The latter obey approximately the PCAC 
relations
\beqn
\label{eq_dnp}
     (m_u + m_d)\langle\bar{u}u + \bar{d}d\rangle
       & \simeq & - 2 f_\pi^2 m_\pi^2~, \nonumber \\
     m_s\langle\bar{s}s\rangle 
       & \simeq & - f_\pi^2 m_K^2~,
\eeqn
where $f_\pi=(92.4\pm0.26)$\MeVE~\cite{pdg} is the pion decay constant.
The complete dimension $D=6$ and $D=8$ operator are parametrized 
phenomenologically using the vacuum expectation values  
$\langle{\cal O}_6\rangle$ and $\langle{\cal O}_8\rangle$, respectively.
Note that in zeroth order $\alpha_s$, \ie, neglecting running quark masses,
nonperturbative dimensions do not contribute to the integral in 
Eq.~(\ref{eq_radler}). Thus in the formula presented in Eq.~(\ref{eq_dnp})
only the gluon and quark condensates contribute to $R$ \via\ the 
logarithmic $s$-dependence of the terms in f\/irst order $\alpha_s$.
\vs
The total Adler $D$-function then reads as the sum of perturbative,
mass and nonperturbative contributions:
\beq
\label{eq_dtot}
      D(s) = D_{\rm P}(s) + D_{\rm mass}(s) + D_{\rm NP}(s)~.
\eeq

%
% -------------------------- Theoretical uncertainties -----------------
%
\subsection*{\it Uncertainties of the Theoretical Prediction}
\label{sec_theory}

Looking at Eq.~(\ref{eq_dtot}) it is instructive to subdivide the
discussion of theoretical uncertainties into three classes:
\begin{itemize}
  \item[(\it i)] {\it The perturbative prediction.} The estimation
                of theoretical errors of the perturbative series is
                strongly linked to its truncation at f\/inite order
                in $\alpha_s$. Due to the incomplete resummation of higher
                order terms, a non-vanishing dependence on the choice
                of the renormalization scheme (RS) and the 
                renormalization scale is left. Furthermore, one has
                to worry whether the missing four-loop order 
                contribution $d_3(\alpha_s/\pi)^4$ gives rise to
                large corrections to the series~(\ref{eq_pertd}).
                On the other hand, these are problems to which 
                any measurement of the strong coupling constant 
                is confronted with, while their impact decreases with
                increasing energy scale. The error on the input parameter 
                $\alpha_s$ itself therefore ref\/lects to some
                extent the theoretical uncertainty of the perturbative
                expansion in powers of $\alpha_s$. \par
                \hspace{0.4cm} 
                Let us use the following, intrinsically dif\/ferent 
                $\alpha_s$ determinations to benchmark our choice of its 
                value and uncertainty. A very robust $\alpha_s$ measurement
                is obtained from the global electroweak f\/it performed
                at the Z-boson mass where uncertainties from perturbative
                QCD are rather small. The value found is 
                $\alpha_s(M_{\rm Z}^2)=0.120\pm0.003$~\cite{jerusalem}.
                A second precise $\alpha_s$ measurement is obtained 
                from the f\/it of the OPE to the hadronic
                width of the $\tau$ and to spectral moments~\cite{aleph_asn}.
                The measurement is dominated by theoretical uncertainties,
                from perturbative origin. In order to ensure the reliability
                of the result, \ie, the applicability of QCD at the $\tau$
                mass scale, spectral moments were f\/itted analogously to
                the analysis presented in the following section. The 
                nonperturbative contribution was found to be lower than $1\%$.
                Additional tests in which the mass scale was reduced down
                to 1~GeV proved the excellent stability of the $\alpha_s$
                determination.
                The value reported by the ALEPH Collaboration~\cite{aleph_asn}
                is $\alpha_s(M_{\rm Z}^2)=0.1202\pm0.0026$. A third, again
                dif\/ferent approach is employed when using lattice
                calculations f\/ixed at $b\bar{b}$ states to adjust $\alpha_s$. 
                The value given in 
                Ref.~\cite{flynn} is $\alpha_s(M_{\rm Z}^2)=0.117\pm0.003$.\par
                \hspace{0.4cm}
                The consistency of the above values using quite dif\/ferent 
                approaches at various mass scales is remarkable and supports 
                QCD as the theory of strong interactions. To be conservative, 
                we choose $\alpha_s(M_{\rm Z}^2)=0.1200\pm0.0045$ as central 
                value for the evaluation of the perturbative contribution to
                the Adler $D$-function. \par
                \hspace{0.4cm}
                Even if it is in principle contained in the uncertainty of
                $\Delta\alpha_s(M_{\rm Z}^2)=0.0045$, we furthermore add
                the total dif\/ference between the results obtained using
                \FOPTCI\ and those from FOPT as systematic error.
  \item[(\it ii)] {\it The quark mass correction.} Since a theoretical 
                evaluation of the integral~(\ref{eq_integral2}) is only 
                applied far from quark thresholds, quark mass corrections
                $D_{\rm mass}$ are small. Without loss of precision, we
                take half of the total correction as systematic uncertainty,
                \ie, we add $D_{\rm mass}\pm D_{\rm mass}/2$ in 
                Eq.~(\ref{eq_dtot}).
  \item[(\it iii)] {\it The nonperturbative contribution}. In order to
                detach the measurement from theoretical constraints
                on the nonperturbative parameters of the OPE, we f\/it
                the dominant dimension $D=4,6,8$ terms by means of
                weighted integrals over the total \ee\ low energy cross section.
                Again, without loss of precision, we take the whole 
                nonperturbative correction as systematic uncertainty, 
                \ie, we add $D_{\rm NP}\pm D_{\rm NP}$ in Eq.~(\ref{eq_dtot}).
\end{itemize}
Another sources of tiny uncertainties included are the errors on the Z-boson
and the top quark masses.

%
% -------------------------- Spectral Moments --------------------------
%
\section{ Spectral Moments}

Constraints on the nonperturbative contributions to $R(s)$
from theory alone are scarce. It is therefore advisable to benef\/it
from the information provided by the explicit shape of the hadronic
width as a function of $s$ in order to determine the magnitude of the
OPE power terms at low energy. We consequently def\/ine the following
spectral moments
\beq
\label{eq_mom}
      R^{kl}(s_0) = \intl_{4m_\pi^2}^{s_0}\frac{ds}{s_0}
                        \left(1-\frac{s}{s_0}\right)^{\!\!k}
                        \left(\frac{s}{s_0}\right)^{\!\!l}R(s)~,
\eeq
where the factor $(1-s/s_0)^k$ squeezes the integrand at the 
crossing of the positive real axis where the validity of the OPE 
is questioned. Its counterpart $(s/s_0)^l$ projects on higher 
energies. The new spectral information is used to f\/it 
simultaneously the phenomenological operators
$\langle(\alpha_s/\pi)GG\rangle$, $\langle{\cal O}_6\rangle$ 
and $\langle{\cal O}_8\rangle$, a procedure which requires 
at least 4 --- better 5 --- input variables considering 
the intrinsic strong correlations between the moments which are 
reinforced by the experimental correlations and by the correlations 
from theoretical uncertainties.
\vs
To predict theoretically the moments, one uses the virtue of Cauchy's 
theorem and the analyticity of the correlator $\Pi(s)$, since a 
direct evaluation of the integral~(\ref{eq_mom}) in the framework of 
perturbation theory (and even OPE) is not possible. With the 
relation~(\ref{eq_rimpi}), Eq.~(\ref{eq_mom}) becomes
\beq
     R^{kl}(s_0) = 6\pi i 
                   \hm\ointl_{|s|=s_0}\hm\hm\frac{ds}{s_0}
                        \left(1-\frac{s}{s_0}\right)^{\!\!k}
                        \left(\frac{s}{s_0}\right)^{\!\!l}\Pi(s)~,
\eeq
and with the def\/inition~(\ref{eq_adler}) of the Adler $D$-function
one further obtains after integration by parts
\beq
\label{eq_momd}
     R^{kl}(s_0) = -\frac{1}{2\pi i}
                   \hm\ointl_{|s|=s_0}\hm\hm\frac{ds}{s}
                    \left[ \frac{k!\,l!}{(k+l+1)!}
                           - \left(\frac{s}{s_0}\right)^{\!\!l+1}
                             \intl_0^1dt\,t^l\left(1-\frac{s}{s_0}t\right)^k
                    \right] D(s)~,
\eeq
where $D(s)$ is obtained from Eq.~(\ref{eq_dtot}).

%
% -------------------------- Measurements of moments --------------------
%
\section{ Data analysis and Determination of the Moments}

Due to the suppression of nonperturbative contributions in 
powers of the energy scale $s$, the critical domain where 
nonperturbative ef\/fects may give residual contributions to
$R(s)$ is the low-energy regime with three active flavours. We
thus choose the energy scale $s_0$ of the f\/it equal to the 
energy scale, where the theoretical evaluation of $R(s)$ shall
start. As demonstrated in isovector vector $\tau$ decays~\cite{aleph_asn}, 
the scale $\sqrt{s}=M_\tau$ is an appropriate scale where 
nonperturbative ef\/fects are present, but essentially controlled 
by the OPE. In our case we manipulate isovector and isoscalar
vector hadronic f\/inal states, \ie, more inclusive data, and 
might expect smaller nonperturbative contributions. We
therefore set the energy cut to $\sqrt{s_0}=1.8~{\rm GeV}$. 
Up to this energy, $R(s)$ is obtained from the sum of the 
hadronic cross sections exclusively measured in the occurring 
f\/inal states. 
\vs
The data analysis follows exactly the line of Ref.~\cite{g_2pap}, 
In addition to the \ee\ annihilation data we use spectral functions 
from $\tau$ decays into two- and four f\/inal 
state pions measured by the ALEPH Collaboration~\cite{aleph_vsf}. 
Extensive studies have been performed in Ref.~\cite{g_2pap}
in order to bound unmeasured modes, such as some ${\rm K\bar K}$ 
or the $\pi^+\pi^-4\pi^0$ f\/inal states, \via\ isospin constraints. 
We bring attention to the straightforward and statistically 
well-def\/ined averaging procedure and error propagation used 
in this paper as in the preceeding one, which takes into account 
full systematic correlations between the cross section measurements. 
All technical details concerning the data analysis and the integration
method used are found in Ref.~\cite{g_2pap}.
\vs
The experimental determination of the spectral moments~(\ref{eq_mom})
is performed as the sum over the respective moments of all exclusively
measured \ee\ f\/inal states (completed by $\tau$ data). We chose
the moments $k=2$, $l=0,\cdots,4$, in order to collect suf\/f\/icient 
information to f\/it the three nonperturbative degrees of freedom.
Neglecting the $s$-dependence of the Wilson coef\/f\/icients in
Eq.~(\ref{eq_ope}), the respective nonperturbative power terms
contribute to the following moments: the dimension $D=4$ term
contributes to $l=0,1$, the $D=6$ term to $l=0,1,2$ and the $D=8$
term contributes to the $l=0,1,2,3$ moments. The $l=4$ moment
receives no direct contribution from any of the considered power 
terms. However its use is not obsolete, since it constrains the
power terms through its correlations to the other moments.
Table~\ref{tab_mom} shows the measured moments together with their
(statistical and systematic) experimental and theoretical errors. 
Additionally given is the sum of the experimental and theoretical 
correlation matrix as it is used in the f\/it.
\vs
\begin{table}[t]
\setlength{\tabcolsep}{1pc}
\begin{center}
{\normalsize
\begin{tabular}{cccc} \hline 
Moments $(k,l)$  
       &  Data   & $\Delta({\rm stat.+sys.})$ 
                           & $\Delta({\rm theo.})$ \\ \hline
(2,0)  & $0.770$ & $0.013$ & $0.024$ \\
(2,1)  & $0.204$ & $0.005$ & $0.012$ \\
(2,2)  & $0.073$ & $0.003$ & $0.001$ \\
(2,3)  & $0.035$ & $0.002$ & $0.001$ \\
(2,4)  & $0.020$ & $0.001$ & $0.001$ \\ \hline 
\end{tabular}
}
\end{center}
\vspace{0.5cm}
\begin{center}
{\normalsize
\begin{tabular}{ c|ccccc } \hline 
 Moments $(k,l)$  
      & (1,0)& (1,1)& (1,2)& (1,3)& (1,4) \\
\hline
(1,0) &  1   & 0.98 & 0.75 & 0.60 & 0.49  \\
(1,1) &  --  &  1   & 0.79 & 0.67 & 0.57  \\
(1,2) &  --  &  --  &  1   & 0.97 & 0.91  \\
(1,3) &  --  &  --  &  --  &  1   & 0.99  \\
(1,4) &  --  &  --  &  --  &  --  &  1    \\
\hline 
\end{tabular}
}
\end{center}
\caption[.]{\label{tab_mom}\it
            Spectral moments measured at $\sqrt{s_0}=1.8~{\rm GeV}$
            with experimental and theoretical errors. Below, the 
            corresponding  correlation matrix containing the quadratic 
            sum of experimental and theoretical covariances.}
\end{table}
Using as input parameters $\alpha_s(M_{\rm Z}^2)=0.1200\pm0.0045$, 
yielding for three flavours $\Lambda_\MSbm^{(3)}=(372\pm76)~{\rm MeV})$, 
and for the mass of the strange quark\footnote
{
   The mass of the strange quark used in this analysis takes one's 
   bearings from the recent experimental determination using the 
   hadronic width of $\tau$ decays into strange f\/inal states,
   $R_{\tau,S}$, reported by the ALEPH collaboration~\cite{shaomin} to
   be $m_s(1~{\rm GeV})=235^{+\,35}_{-\,42}~{\rm MeV}$. The error on
   this mass has no inf\/luence on the present analysis since, conservatively, 
   $50\%$ of the total mass contribution given in Eq.~(\ref{eq_dmass}) 
   is taken as corresponding systematic uncertainty.
} 
$m_s(1~{\rm GeV})=0.230~{\rm GeV}$, while setting $m_u=m_d=0$, the 
adjusted nonperturbative parameters are
\beqn
\label{eq_nonp}
     \left\langle\frac{\alpha_s}{\pi} GG\right\rangle
                        &=&  (0.037\pm0.019)~{\rm GeV}^4~, \nonumber\\
     \langle O_6\rangle &=& -(0.002\pm0.003)~{\rm GeV}^6~, \nonumber\\
     \langle O_8\rangle &=&  (0.002\pm0.003)~{\rm GeV}^8~,
\eeqn
with $\chi^2/{\rm d.o.f}=1.5/2$. The correlation coef\/f\/icients 
between the f\/itted parameters are 
$\rho(\langle(\alpha_s/\pi)GG\rangle,\langle O_6\rangle)=-0.41$,
$\rho(\langle(\alpha_s/\pi)GG\rangle,\langle O_8\rangle)=0.55$
and $\rho(\langle O_6\rangle,\langle O_8\rangle)=-0.98$. As a
test of stability we have additionally f\/itted the $k=2,l=0,\dots,4$
spectral moments at $\sqrt{s_0}=2.1$. The dif\/ference between
the results of this f\/it and Eq.~(\ref{eq_nonp}) is included
into the parameter errors given in Eq.~(\ref{eq_nonp}). With these
values, the corrections to $R(s_0)$ at $\sqrt{s_0}=1.8~{\rm GeV}$
amount to $0.16\%$ from the strange quark mass and $<0.1\%$ from the
nonperturbative power terms. The gluon condensate can be compared 
to the standard value obtained from charmonium sum rules, 
$\left\langle(\alpha_s/\pi)GG\right\rangle=(0.017\pm0.004)
~{\rm GeV}^4$~\cite{reinders}, which lies below our value. 
However, another estimation~\cite{bertlmann} using f\/inite 
energy sum rule techniques on \ee\ data gives the value of 
$\left\langle(\alpha_s/\pi)GG\right\rangle
=(0.044^{\;+0.004}_{\;-0.021})~{\rm GeV}^4$ in agreement with the 
result~(\ref{eq_nonp}). F\/itting the moments when f\/ixing the dimension
$D=6,8$ contributions reduces the gluon condensate to $0.010\pm0.002$.
One may additionally compare the f\/itted dimension $D=6,8$ operator 
to the results obtained from the $\tau$ vector spectral functions, 
keeping in mind that only the isovector amplitude contributes
in this case and thus the more inclusive isoscalar plus isovector
moments from \ee\ annihilation are expected to receive
smaller nonperturbative contributions\footnote
{
   The vacuum saturation hypothesis of\/fers a relation between the 
   dimension $D=6$ contribution and the light quark condensates~\cite{svz}. 
   Using the formulae~(\ref{eq_nonpdef}) and (\ref{eq_dnp}) one has
   \beqns
      \langle{\cal O}_6\rangle 
         \simeq
           - \frac{224}{81}\pi\alpha_s(s_0)\langle\bar{q}q\rangle^2
         \approx - 10^{-3}~{\rm GeV}^6~.
   \eeqns
}. 
Reexpressing the results of Ref.~\cite{aleph_asn} in terms of the 
def\/inition adopted in Eq.~(\ref{eq_nonpdef}) we obtain 
$\langle O_6\rangle_{I=1}=-(0.0042\pm0.0006)~{\rm GeV}^6$ and 
$\langle O_8\rangle_{I=1}=(0.0062\pm0.0007)~{\rm GeV}^8$.
\vs
As a cross check of the spectral moment analysis, we f\/it
the three nonperturbative parameters and $\alpha_s$. The theoretical 
error applied reduces essentially to the theoretical uncertainties of
the QCD perturbative series estimated in Ref.~\cite{aleph_asn}
to be $\Delta\alpha_s(M_{\rm Z}^2)=0.0023$ at the $\tau$ mass scale 
which is of the same magnitude as the scale $\sqrt{s_0}=1.8~{\rm GeV}$ 
used here. The f\/it of the $k=2,l=0,\dots,4$ moments yields 
$\alpha_s(M_{\rm Z}^2)=0.1205\pm0.0053$ ($\chi^2/{\rm d.o.f}=1.4/1$), 
in agreement with the values from other analyses cited above.
The values for the gluon condensate and the higher dimension 
contributions are consistent with those of Eq~(\ref{eq_nonp}).
F\/itting \assz\ and the gluon condensate at $\sqrt{s_0}=2.1~{\rm GeV}$
when f\/ixing the dimension $D=6,8$ contributions at the values
of Eq.~(\ref{eq_nonp}) yields $\alpha_s(M_{\rm Z}^2)=0.121\pm0.011$
and $\left\langle(\alpha_s/\pi)GG\right\rangle=(0.042\pm0.003)~{\rm GeV}$.
The consistency of these results with those obtained at 
$\sqrt{s_0}=1.8~{\rm GeV}$ supports the stability of the OPE approach 
within the energy regime where it is applied. Varying the c.m. energy
$s_0$ and the weights $k,l$ of the moments used to f\/it the 
nonperturbative contributions gives a measure of the compatibility 
between the data and the OPE approach. Dif\/ferences found between 
the f\/itted parameters are included as systematic uncertainties 
in the errors of the values~(\ref{eq_nonp}).

%
% -------------------------- The valuation of the integral --------------
%
\section{ Evaluation of \daqedhZ\ and \amuhad}

In order to get the most reliable central value of the perturbative 
$R(s)$ prediction which enters the integrals~(\ref{eq_integral2})
and (\ref{eq_integralamu}), we use the whole set of formulae given 
in Ref.~\cite{kuhn1}, including mass corrections up to order 
$\alpha_s^2$. Despite this theoretical precision, the uncertainties 
keep conservatively estimated as described in Section~\ref{sec_theory}.
\vs
Since deeply nonperturbative phenomena are not predictable within
the OPE approach, we use experimental data to cover energy regions
near quark thresholds. The low energy results for \daqedhZ\ and
\amuhad\ from Ref.~\cite{g_2pap} including $\tau$ data are taken for 
$\sqrt{s}\le1.8~{\rm GeV}$. The narrow $\omega$, $\phi$, $J/\psi$ 
and $\Upsilon$ resonances are parametrized using relativistic 
Breit-Wigner formulae as described in Refs.~\cite{eidelman,g_2pap}.
In addition, $R$ measurements of the continuum contributions in the 
environment of the $c\bar{c}$ threshold are taken from the experiments 
specified in Ref.~\cite{g_2pap}. The technical aspects of the integration 
over data points are also discussed in Ref.~\cite{g_2pap}. No continuum 
data are available at $b\bar{b}$ threshold energies in the range of 
$10.6~{\rm GeV}\le E\le12~{\rm GeV}$. A recent analysis of the 
$\Upsilon$ system~\cite{upsilon} however showed that essentially 
within the perturbative approach of Ref.~\cite{kuhn1} it is 
possible to predict weighted integrals over the $b\bar{b}$ states.
The uncertainty of this approach corresponds to an estimated error 
of $\Delta\alpha_s(M_{\rm Z}^2)=0.008$. We therefore use this uncertainty 
of $\alpha_s$ for the QCD prediction at $b\bar{b}$ threshold energies.
Included is a small uncertainty originating from scale ambiguities when
matching the ef\/fective theories of four and f\/ive f\/lavours.
\vs
For the theoretical evaluation of the integrals~(\ref{eq_integral2}),
(\ref{eq_integralamu}) (\via\ Eq.~(\ref{eq_radler})), we use the 
following variable settings:
\beqns
   M_{\rm Z}                 &=& (91.1867\pm0.0020)~{\rm GeV}~\cite{jerusalem}, \\
   \alpha(0)                 &=& 1/137.036~, \\
   \alpha_s(M_{\rm Z}^2)     &=& 0.1200 \pm 0.0045~, \\[0.2cm]
   m_u = m_d                 &\equiv& 0~, \\
   m_s(1~{\rm GeV})          &=& 0.230~{\rm GeV}~, \\
   m_c(m_c)                  &=& 1.3~{\rm GeV}~, \\
   m_b(m_b)                  &=& 4.1~{\rm GeV}~, \\
   m_t(m_t)                  &=& (175.6\pm5.5)~{\rm GeV}~\cite{jerusalem}~,
\eeqns  
and the values~(\ref{eq_nonp}) for the nonperturbative contributions.
There are no errors assigned to the light quark masses, since the half 
of the total quark mass correction is taken as systematic uncertainty.
The error on $m_t$ is needed in order to estimate the systematic 
uncertainty of the matching scale when turning from f\/ive to six 
f\/lavours. Table~\ref{tab_alphares} shows the experimental and 
theoretical evaluations of \daqedhZ\ and \amuhad\ for the respective 
energy regimes. The upper star denotes the values used for the f\/inal 
summation of \daqedhZ\ given in the last line. At $E\simeq 3.8~{\rm GeV}$ 
a non-resonant $D\bar{D}$ production might contribute to the 
continuum\footnote
{
   Such a contribution must be tiny since it is 
   suppressed by its form factor and the $(1-4M_D^2/s)^{3/2}$
   threshold behaviour of a pair of spin 0 particles.
}. 
We therefore use experimental data to cover energies from 3.7--5.0~GeV. 
\vs
\begin{table}[t]
\setlength{\tabcolsep}{0.7pc}
\begin{center}
{\normalsize
\begin{tabular}{c|ccc|c} \hline &&& \\
$E_{\rm min}$ -- $E_{\rm max}$~(GeV) 
                         & 
          \mc{2}{c}{\rs{$\Delta \alpha_{\rm had}(M_{\rm Z}^2)\times10^{4}$}} 
  & $\sigma$ & $a_\mu^{\rm had}\times10^{10}$ \\
             & \rs{Data }          & \rs{Theory}           & 
  & \\ 
\hline &&& \\
\rs{$4m_\pi^2$ -- 1.8} & \rs{$56.9\pm1.1^{(*)}$} &    \mc{1}{c}{\rs{--}} & \rs{--} 
  & \rs{$636.49\pm7.41$ (D)} \\
1.8 -- 3.700           &  $32.4\pm3.1~~~$        &  $24.50\pm0.33^{(*)}$ & 2.5 
  & $33.84\pm0.53$ (T)    \\
&&& \\
\rs{$\psi$(3770)}        &\rs{$ 0.29\pm0.08^{(*)}$}&  $17.06\pm0.58~~~$    & 0.5 
  &  \rs{$0.17\pm0.02$ (D)} \\
\rs{3.700 -- 5.000}      & \rs{$15.8\pm1.7^{(*)}$} &                       &  
  &  \rs{$6.93\pm0.62$ (D)} \\
5.000 -- 10.500          &  $39.9\pm1.4~~~$        &  $41.55\pm0.40^{(*)}$ & 1.0 
  & $7.43\pm0.09$ (T) \\
&&& \\
\rs{$\Upsilon$(4S,10860,11020)}
                         &\rs{$ 0.38\pm0.10~~~$}&  $ 8.19\pm0.32^{(*)}$    & 0.4  
  & $0.55\pm0.03$ (T) \\
\rs{10.500 -- 12.000}    &  \rs{$ 7.6\pm0.5~~~$}   &                       &      \\
12.000 -- 40.000         &  $75.2\pm2.7~~~$        &  $77.96\pm0.30^{(*)}$ & 1.0  
  & $1.64\pm0.02$ (T) \\
&&& \\
\rs{40.000 -- $\infty$}  &     \mc{1}{c}{\rs{--}}  &\rs{$41.98\pm0.22^{(*)}$}&\rs{--}
  & \rs{$0.16\pm0.00$ (T)} \\
$J/\psi$(1S,2S) &  $ 9.68\pm0.68^{(*)}$ & -- & -- 
  & $7.80\pm0.46$ (D) \\
 &&& \\
\rs{$\Upsilon$(1S,2S,3S)}&\rs{$ 0.98\pm0.15^{(*)}$}& \rs{--}               & \rs{--}
  & \rs{$0.09\pm0.01$ (D)} \\ 
\hline
 &&& \\ 
\rs{$4m_\pi^2$ -- $\infty$} & 
             \mc{3}{c|}{ \rs{$277.8\pm2.2_{\rm exp}\pm1.4_{\rm theo}$ }} 
  & \rs{$695.1\pm7.5_{\rm exp}\pm0.7_{\rm theo}$} \\
\hline
\end{tabular}
}
{\footnotesize 
\parbox{12cm}
{
\vspace{0.2cm}
$^{(*)}\,$Value used for f\/inal result of \daqedhZ\ (last line).
}}
\end{center}
\caption[.]{\label{tab_alphares}\it
            Contributions to \daqedhZ\ and to \amuhad\ from the 
            different energy regions. The ``$\sigma$'' column gives the 
            standard deviation between the experimental and theoretical 
            evaluations of \daqedhZ. ``(D)/(T)'' in the last column 
            stands for evaluation using Data/Theory.}
\end{table} 
A $20\%$ correlation is assumed between the analytic evaluations
of the narrow resonances where in each case a Breit-Wigner 
formula is applied. The theoretical errors are by far dominated by
uncertainties from $\alpha_s$ and the dif\/ference FOPT/\FOPTCI.
For instance, the f\/irst energy interval where theory is applied
($E\in\{1.8-3.7~{\rm GeV}\}$) receives the error contributions 
\beqns
      \Delta \alpha_{\rm had}(M_{\rm Z}^2)\times10^4 
        = 24.502\pm0.332\;
           {\small\left\{ \begin{array}{rcl} 0.243 &-&  \Delta\alpha_s \\
                                      0.223 &-&  {\rm perturbative~prediction} \\
                                      0.036 &-&  {\rm quark~mass~correction} \\
                                      0.003 &-&  {\rm nonperturbative~parts} \\
                                     <0.001 &-&  \Delta M_{\rm Z} 
                   \end{array}
            \right.}
\eeqns
where the small contributions from quark masses and nonperturbative
dimensions show that the perturbative QCD calculation is very solid 
here. Remember that only ${\rm ln}\,s$-dependent nonperturbative terms
contribute to $R(s)$. Theoretical errors of dif\/ferent energy 
regions are added linearily except the uncertainties at $b\bar{b}$
threshold that are (partly) from individual origin so that a
$50\%$ correlation to other energy regimes is estimated here.
Looking at Table~\ref{tab_alphares} one notices the remarkable 
agreement between experimental data and theoretical predictions
of \daqedhZ\ even in the $c\bar{c}$ quark threshold regions where 
strong oscillations occur. The experimental results of $R(s)$ 
and the theoretical prediction are shown in F\/ig.~\ref{fig_data}. 
The shaded bands depict the regions where data are used instead 
of theory to evaluate the respective integrals. Good agreement 
between data and QCD is found above 8~GeV, while at lower energies 
systematic deviations are observed. The $R$ measurements in this 
region are essentially provided by the $\gamma\gamma2$~\cite{E_78} 
and MARK~I~\cite{E_96} collaborations. MARK~I data above 5~GeV lie 
systematically above the measurements of the Crystal Ball~\cite{CB} 
and MD1~\cite{MD1} Collaborations as well as the QCD prediction.
\vs
\begin{figure}[p]
\epsfxsize17cm
\centerline{\epsffile{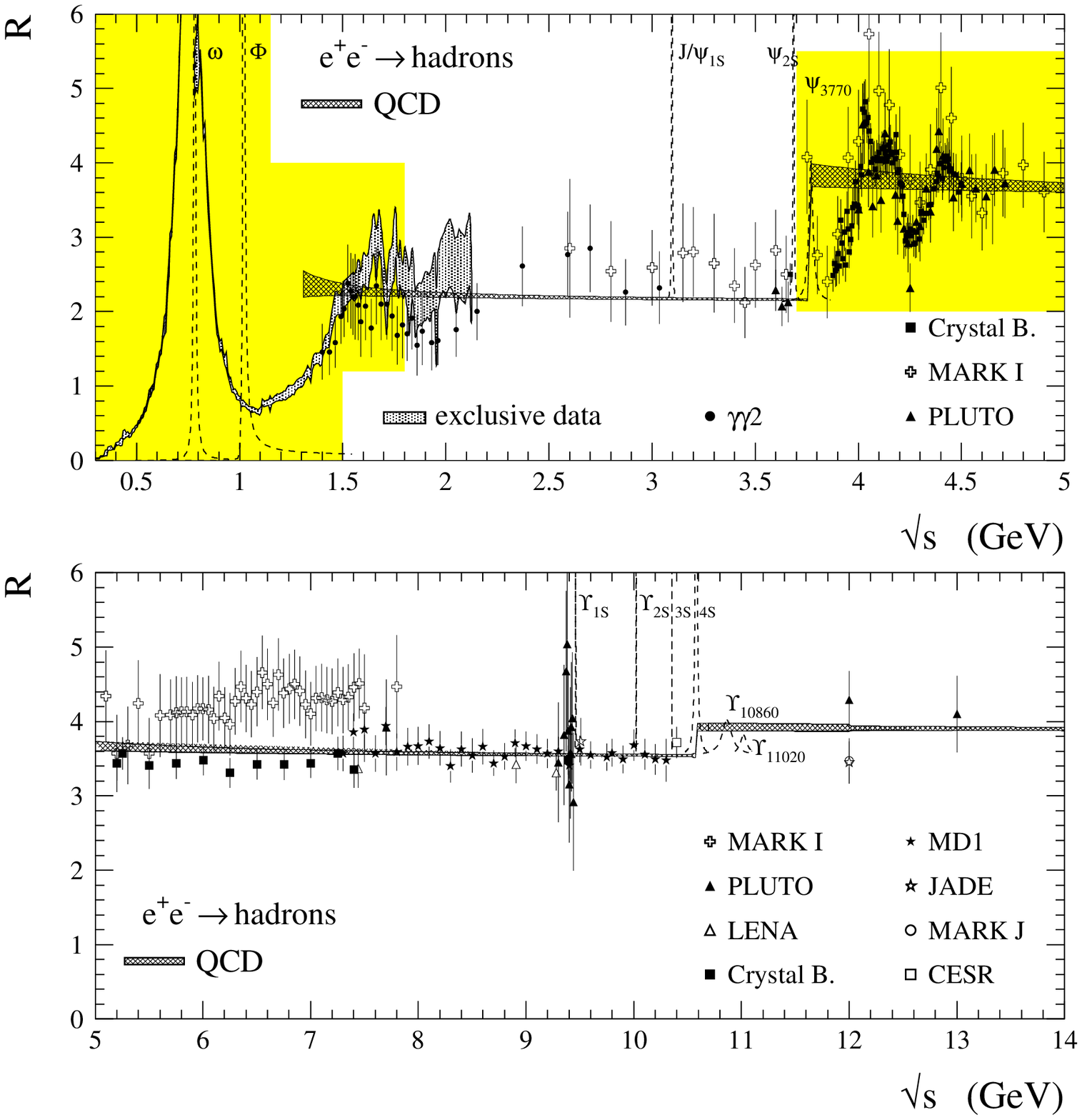}}
\vspace{-0.2cm}
\caption[.]{\label{fig_data}\it 
            Inclusive hadronic cross section ratio in \ee\
            annihilation versus the c.m. energy $\sqrt{s}$. 
            Additionally shown is the QCD prediction of the continuum 
            contribution as explained in the text. The shaded areas 
            depict regions were experimental data are used for the 
            evaluation of \daqedhZ\ and \amuhad\ in addition to the 
            measured narrow resonance parameters. The exclusive 
            \ee\ cross section measurements at low c.m. energies
            are taken from DM1,DM2,M2N,M3N,OLYA,CMD,ND and
            $\tau$ data from ALEPH (see Ref.~\cite{g_2pap}
            for detailed information).}
\end{figure}

%
% ----------------------- Results ---------------------------------------
%
\section{Results}

According to Table~\ref{tab_alphares}, the combination of the theoretical
and experimental evaluations of the integrals~(\ref{eq_integral2}) 
and (\ref{eq_integralamu}) yield the f\/inal results
\beq
\label{eq_asres}
\begin{array}{|rcl|}
  \hline 
  & & \\
   ~~~\Delta\alpha_{\rm had}(M_{\rm Z}^2) 
      &=& (277.8 \pm 2.2_{\rm exp} \pm 1.4_{\rm theo})\times10^{-4}~~~ \\
  & & \\
   ~~~\alpha^{-1}(M_{\rm Z}^2) 
      &=& 128.923 \pm 0.030_{\rm exp} \pm 0.019_{\rm theo}~~~ \\ 
  & & \\
   ~~~a_\mu^{\rm had}
      &=& (695.1 \pm 7.5_{\rm exp} \pm 0.7_{\rm theo})\times10^{-10}~~~ \\
  & & \\
   ~~~a_\mu^{\rm SM}
      &=& (11\,659\,164.6 \pm 7.5_{\rm exp} \pm 4.1_{\rm theo})\times10^{-10}~~~ \\
  & & \\ 
  \hline
\end{array}
\eeq
The total $a_\mu^{\rm SM}$ value includes an additional contribution
from non-leading order hadronic vacuum polarization summarized in 
Refs.~\cite{krause2,g_2pap} to be 
$a_\mu^{\rm had}[(\alpha/\pi)^3]\,=\,(-16.2 \,\pm\,4.0)\times10^{-10}$,
where the error originates essentially from the uncertainty on the
theoretical evaluation of the light-by-light scattering type of 
diagrams~\cite{light1,light2}.
\begin{figure}[t]
\epsfxsize17cm
\centerline{\epsffile{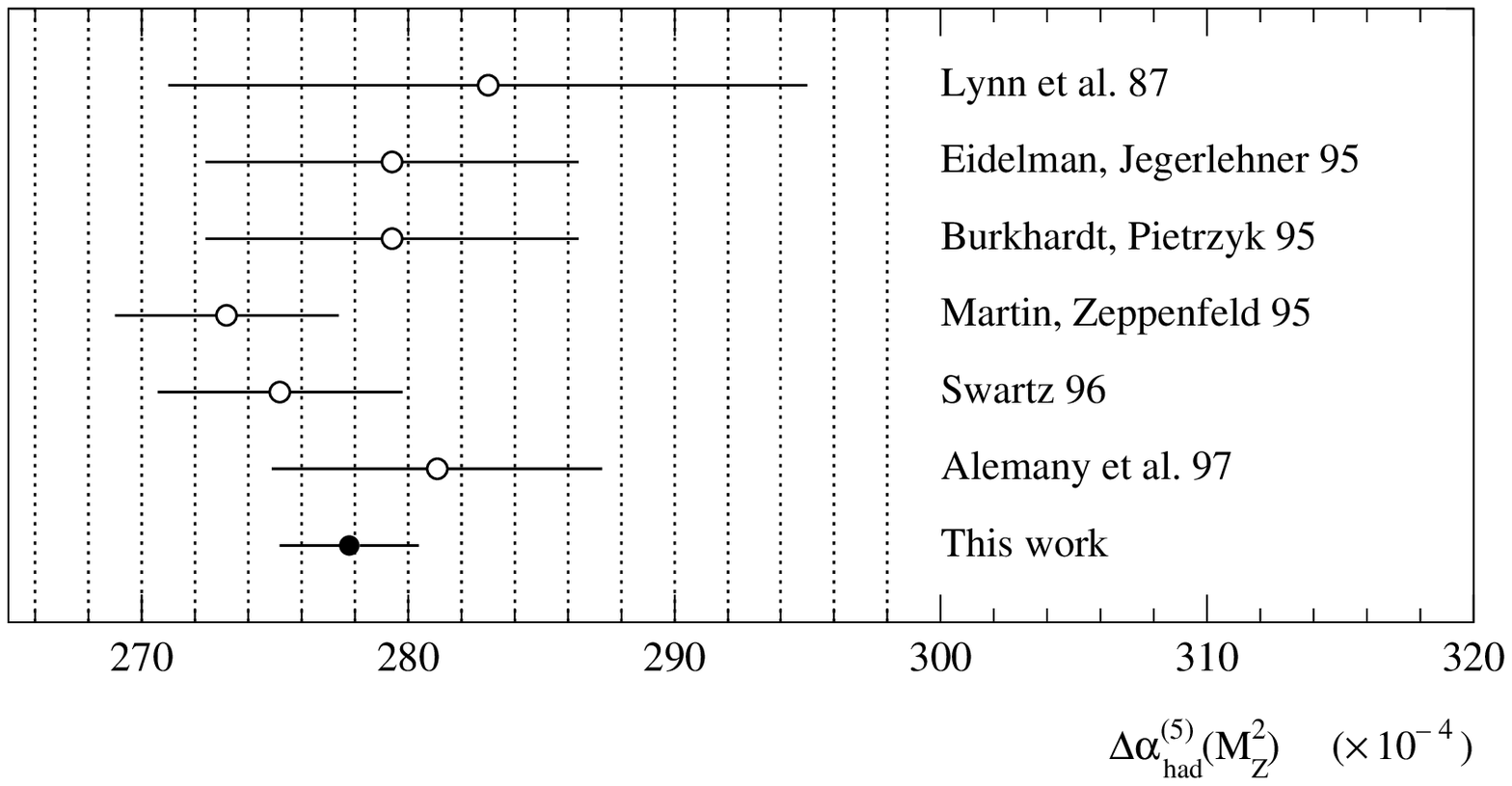}}
\caption[.]{\label{fig_results}\it 
            Comparison of estimates of and \daqedhZ. The values are
            taken from 
            Refs.~\rm\cite{lynn,eidelman,burkhardt,martin,swartz,g_2pap}.}
\end{figure}
F\/ig.~\ref{fig_results} shows a compilation of published results
for the hadronic contribution \daqedhZ. Some authors give the 
hadronic contribution for the f\/ive light quarks only and add 
the top quark contribution of 
$\Delta\alpha_{\rm top}(M_{\rm Z}^2)=-0.6\times10^{-4}$ 
separately. This has been corrected for in the f\/igure. The present 
inclusion of theoretical evaluations yields an improvement in the 
\daqedhZ\ precision of a factor of two. The improvement of the
precision on \amuhad\ with respective to the analysis presented
in Ref.~\cite{g_2pap} amounts to $20\%$.
\vs
The analysis using the spectral moments and also the measurement
of $\alpha_s$ at the $\tau$ mass scale~\cite{aleph_asn} showed
concordantly that the OPE can be safely applied in order to
predict integrals over inclusive spectra at relatively low energy 
scales. Nonperturbative contributions are indeed tiny and well
below the envisaged precision. This approach leads to a large
improvement in precision, as compared to the method used in
Ref.~\cite{martin} where perturbative QCD was assumed to be valid 
only above 3~GeV. Also the value used for \asZ\ in the latter
analysis was less precise than the current value. F\/inally,
better and more complete experimental information, in particular 
at low energy, is used in the present work. Experimental 
data on $R$ are still necessary at very low energies 
and at the $c\bar{c}$ quark production threshold.
Uncontrolled nonperturbative ef\/fects spoil the hadronic spectra
and make the OPE approach unreliable. Thus, the final results on 
both, \daqedhZ\ and \amuhad, are still dominated by experimental 
uncertainties, where in particular the unmeasured $\pi^+\pi^-4\pi^0$ 
and ${\rm K}\bar{\rm K}\pi\pi$ low energy f\/inal states, which must
be conservatively bound using isospin symmetry~\cite{g_2pap}, as
well as the two-pion f\/inal state in the latter case,
contribute with large errors. Also more precise data of
the $c\bar{c}$ continuum at threshold energies are needed.
\vs
We use the improved precision on $\alpha(M_{\rm Z}^2)$ to repeat
the global electroweak f\/it in order to adjust the mass of the
Standard Model Higgs boson, $M_{\mathrm{Higgs}}$. As electroweak
and heavy flavour input parameters we use values measured by the 
LEP, SLD, CDF, D0, CDHS, CHARM and CFFR collaborations, which have
been collected and averaged in Ref.~\cite{jerusalem}. The prediction
of the Standard Model is obtained from the ZFITTER electroweak 
library~\cite{leplib}. The f\/it adjusts $M_{\rm Z}$, $M_{\rm top}$
and $\alpha(M_{\rm Z})$ which are allowed to vary within their errors. 
Freely varying parameters are $\alpha_s$ and $M_{\rm Higgs}$. 
We obtain $\alpha_s(M_{\mathrm Z}^2)\,=\,0.1198\,\pm\,0.0031$ 
in agreement with the experimental value of 0.122\pms0.006~\cite{alphas1}
from the analyses of QCD observables in hadronic Z decays at LEP.
The f\/itted Higgs boson mass is $129^{+103}_{-62}~{\rm GeV}$ with
$\chi^2=16.4/15$, compared to $105^{+112}_{-62}~{\rm GeV}$ when using 
the previous value of \aqedZ\ from Ref.~\cite{g_2pap}. An additional 
error of 50~GeV should be added to account for theoretical 
uncertainties~\cite{leplib}. Doing so, we obtain an upper limit 
for $M_{\mathrm{Higgs}}$ of 398~GeV at 95\pc\ CL.
\begin{figure}[t]
\epsfxsize10cm
\centerline{\epsffile{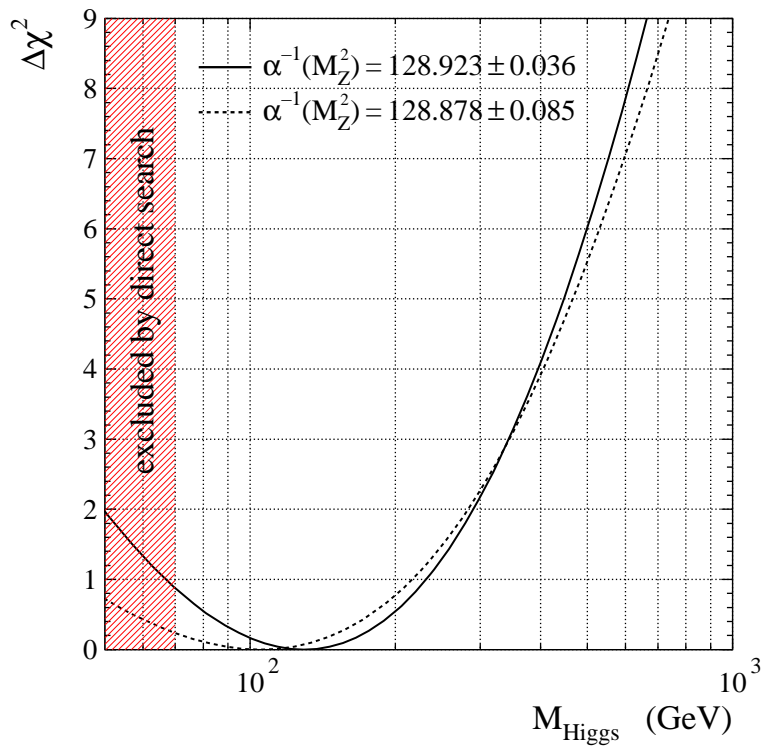}}
\caption[.]{\it Constraint fit results for the previous and the 
             new value of \aqedZ\ as a function of the Higgs mass.}
\label{fig_mhiggs}
\end{figure}

Fig.~\ref{fig_mhiggs} depicts the variation of $\Delta\chi^2$ 
as a function of the Higgs boson mass for the new and previously used 
values of \aqedZ~\cite{g_2pap}.
\vs
\begin{table}[t]
\setlength{\tabcolsep}{1pc}
\begin{center}
{\normalsize
\begin{tabular}{cccc} 
                & \\
                & $\Delta\,{\rm sin}^2\theta_{\rm eff}^{\rm lept}$  \\
                &     \\[-0.4cm] \hline
                &     \\[-0.4cm]
Experiment      & 0.00023                              \\
\aqedZ          & 0.00023 $\stackrel{\rm this~work}{\Longrightarrow}$ 0.00010 \\
$m_t$           & 0.00018                              \\
Theory          & 0.00014                              \\
$M_{\rm Higgs}$ & 0.00160 $[M_{\rm Higgs}=60$--$1000$~GeV$]$
                  $\stackrel{\rm F\/it}{\Longrightarrow}$ 
                  $M_{\rm Higgs}=129^{+103}_{-62}~{\rm GeV}$ \\
                &     \\[-0.4cm] \hline
\end{tabular}
}
\end{center}
\caption[.]{\label{tab_sin}\it
            Dominant uncertainties of input values of the Standard Model
            electroweak fit expressed in terms of 
 $\Delta\,{\rm sin}^2\theta_{\rm eff}^{\rm lept}$~\rm\cite{jerusalem,blondel}.}
\end{table}
Table~\ref{tab_sin} shows the dominant uncertainties of the input 
values of the Standard Model f\/it expressed in terms of 
$\Delta\,{\rm sin}^2\theta_{\rm eff}^{\rm lept}$. This work reduces 
the uncertainty of \aqedZ\ on $M_{\rm Higgs}$ below the uncertainties 
from the experimental value of ${\rm sin}^2\theta_{\rm eff}^{\rm lept}$ 
or from the Standard Model, \ie, theoretical origin.

%
% -------------------------- Conclusions --------------------------------
%

\section{ Conclusions}

We have reevaluated the hadronic vacuum polarization contribution to 
the running of the QED f\/ine structure constant, $\alpha(s)$, at
$s=M_{\rm Z}^2$ and to the anomalous magnetic moment of the muon,
$a_\mu$. We employed perturbative and nonperturbative 
QCD in the framework of the Operator Product Expansion in order
to extend the energy regime where theoretical predictions are 
reliable. In addition to the theory, we used data from \ee\ 
annihilation and $\tau$ decays to cover low energies and quark 
thresholds. The extended theoretical approach reduces the 
uncertainty on \daqedhZ\ by more than a factor of two. Our results are
\daqedhZ\,$=(277.8\pm2.6)\times 10^{-4}$, propagating $\alpha^{-1}(0)$ 
to $\alpha^{-1}(M_{\rm Z}^2)=(128.923\pm0.036)$, and 
$a_\mu^{\rm had}=(695.1\pm7.5)\times10^{10}$ which yields the Standard 
Model prediction $a_\mu^{\rm SM}=(11\,659\,164.6\pm8.5)\times10^{-10}$.
The new value for $\alpha(M_{\rm Z}^2)$ improves the constraint on 
the mass of the Standard Model Higgs boson to
$M_{\rm Higgs}=129^{+103}_{-62}~{\rm GeV}$.

%
% -------------------------- Acknowledgments ----------------------------
%
\section*{\large \it Acknowledgments}
{\small 
We gratefully acknowledge E.~de Rafael for carefully reading the 
manuscript and J.~K\"uhn for helpful discussions.
}

%
% -------------------------- Bibliography -------------------------------
%

\end{document}